\documentclass{jetpl}
\twocolumn

\lat

\title{New Narrow Nucleon N$^*(1685)$}

\rtitle{New Narrow Nucleon N$^*(1685)$}

\sodtitle{New Narrow Nucleon N$^*(1685)$}

\author{V.~Kuznetsov$^{1,2}$, M.V.~Polyakov$^{3,4}$}

\rauthor{V.~Kuznetsov,  M.V.~Polyakov}

\sodauthor{V.~Kuznetsov,  M.V.~Polyakov}

\address{$^1$ Kyungpook National University, 702-701, Daegu, Republic of
Korea\\
$^2$ Institute for Nuclear Research, 117312, Moscow, Russia, \\
$^3$ Institute f\"ur Theoretische Physik II, Ruhr-Universit\"at
Bochum,
D - 44780 Bochum, Germany, \\
$^4$Petersburg  Nuclear Physics Institute, Gatchina, 188300, St.
Petersburg, Russia.}

\abstract{We argue that the existence of a new narrow ($\Gamma
\leq$~25~MeV) nucleon resonance N$^*$(1685) is strongly supported
by recent data on $\eta$ photoproduction off the nucleon. The
resonance has much stronger photo-coupling to the neutron than to
the proton. This nucleon resonance is a good candidate for the
non-strange member of the exotic anti-decouplet of baryons -- the
partner of the pentaquark $\Theta^+$. All up to date known
properties of new N$^*$(1685) are summarized.}


\begin{document}

\maketitle

Secondly, the exotic $\Theta^+$ baryon always must be accompanied
by its siblings. A multiplet containing pentaquarks should also
contain baryons with non-exotic ``3-quark" quantum numbers. The
minimal SU$_{\rm fl}(3)$ multiplet containing pentaquarks is the
anti-decouplet of baryons.  In  the anti-decouplet \cite{dia}
there are two types of baryons with non-exotic quantum numbers:
the isodoublet of non-strange nucleons (N$^*$) and the isotriplet
of $S=-1$ $\Sigma^*$'s. In the Chiral Quark-Soliton model
($\chi$QSM) the spin-parity quantum numbers of the anti-decouplet
members are unambiguously predicted to be $J^P=\frac 12^+$
\cite{dia}, so that the N$^*$ from the anti-decouplet is predicted
to be a $P_{11}$ nucleon resonance. One of the striking properties
of N$^*$ is that it can be excited by an electromagnetic probe
from the neutron target much stronger than from the proton
one~\cite{max}. The photoexcitation of charged isocomponent of
N$^*$ is possible only due to SU$_{\rm fl}(3)$ violation;
therefore, it is suppressed by a factor $\sim 1/10$ in the
amplitude. The mass of the anti-decouplet N$^*$ was predicted
 to be around 1680~MeV in Refs.~\cite{dia1,str}. Its width is
predicted to be in the range of tens of MeV ($\Gamma \leq 30$~MeV)
 with a very small
coupling to the $\pi N$ channel \cite{str}. The preferred decay
channels are predicted to be $\eta N$, $\pi \Delta$ and $K
\Lambda$
~\cite{dia,str,michal1,michal2,michal3,guz2,Ledwig:2008rw}.
\begin{figure}
\vspace*{-0.4cm}
\centerline{\resizebox{0.5\textwidth}{!}{\includegraphics{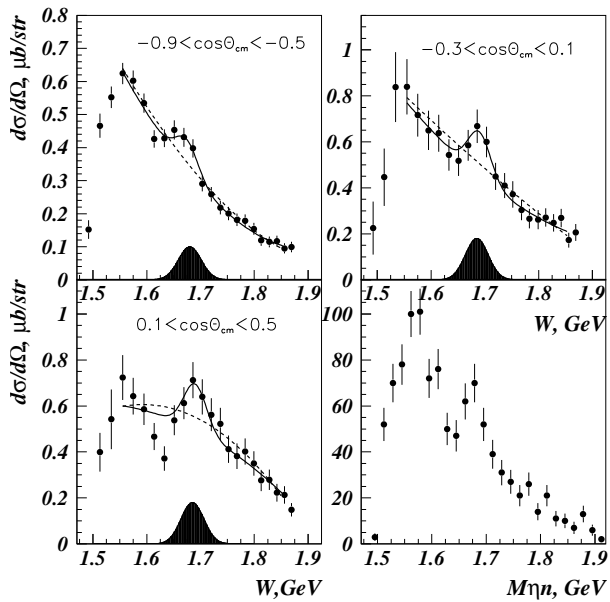}}}
\caption{Fig. 1. Quasi-free cross sections and $\eta n$ invariant
mass spectrum (low right panel) for the $\gamma n \to \eta n$
reaction (data from \protect\cite{gra1}). Solid lines are the fit
by the sum of 3-order polynomial and narrow state. Dashed lines
are the fit by 3-order polynomial only. Dark areas show the
simulated signal of a narrow state. } \label{fig:etan}
\vspace{-0.3cm}
\end{figure}

Predictions of Refs.~\cite{max,dia1,str} encouraged one of the
authors (V.~K.) to push forward the study the $\eta$
photoproduction on the neutron at GRAAL. In 2004 these efforts led
to the observation of a narrow peak in the quasi-free neutron
cross section and in the $\eta n$ invariant mass
spectrum~\cite{Kuznetsov04,gra1}, see Fig.~1.

The original observation of Refs.~\cite{Kuznetsov04,gra1} has been
recently confirmed by two other groups: CBELSA/TAPS~\cite{kru} and
LNS-Sendai~\cite{kas}. In all three experiments an enhancement in
the quasi-free cross-section\footnote{For brevity we call this
enhancement ``neutron anomaly".} on the neutron was found.
Moreover, the GRAAL and CBELSA/TAPS groups have observed narrow
peaks in the $\eta n$ invariant mass spectrum at $1680-1685$~MeV.
The position of the peaks are $\sim 1680$~MeV at GRAAL data (see
low-right panel of Fig.~1) and $\sim 1683$~MeV at CBELSA/TAPS data
(see Fig.~2). The width of the peaks is 40~MeV in the GRAAL data
and $60\pm 20$~MeV in the CBELSA/TAPS data. In both experiments
the width is dominated by the instrumental resolution. We note
that the cross section in the proton channel does not exhibit any
strong enhancement around $W\sim 1680-1685$~MeV (see e.g. Fig.~2).

A simple and concise explanation of the ``neutron anomaly" and the
peak in the $\eta n$ invariant mass is the existence of a new
narrow nucleon resonance with much stronger photocoupling to the
neutron than to the proton predicted in Refs.~\cite{max,dia1,str}.
Due to the weak photocoupling to the proton, the expected signal
of N$^*$ in the proton channel requires high precision and high
resolution of the data.

\begin{figure}
\vspace*{-0.0cm}
\centerline{\resizebox{0.4\textwidth}{!}{\includegraphics{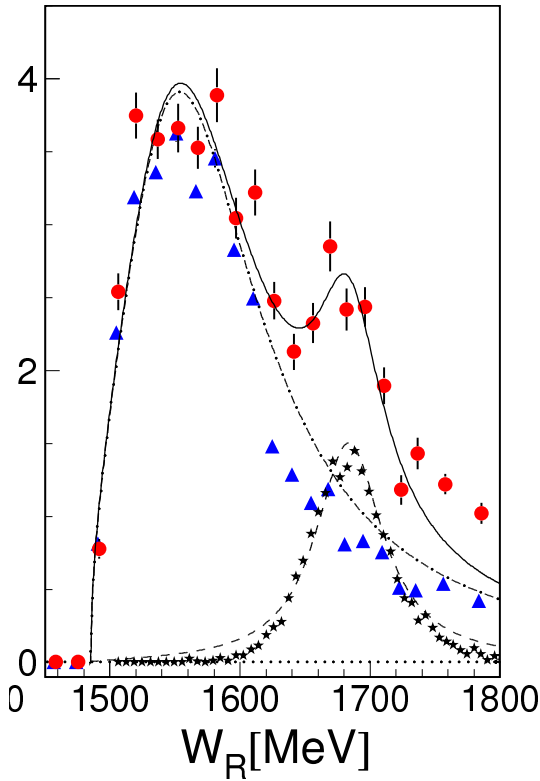}}}
\caption{Fig. 2. $M(\eta n)$ spectrum from CBELSA/TAPS
\protect\cite{kru} (filled circle) in comparison with $M(\eta p)$
spectrum (filled triangles) Stars show the simulated signal of a
narrow state.} \vspace*{-0.3cm} \label{fig:kru}
\end{figure}

Alternative theoretical explanations of the ``neutron anomaly"
were suggested in Refs.~\cite{skl,aniso}. The authors demonstrated
that the bump in the $\gamma n \to \eta n$ cross section could be
explained in terms of photoexcitation  of the
known $S_{11}(1650)$ and $P_{11}(1710)$ (or $S_{11}(1535)$ and
$S_{11}(1650)$) resonances. The authors of Refs.~\cite{skl,aniso}
tuned the neutron photocouplings of these known resonances in
order to obtain a  bump in the $\gamma n \to \eta n$ quasi-free
cross section. The proton photocouplings were not touched. This
implies that  all observables in the proton channel, predicted
by these models, do not possess any irregularities at $W\sim
1680-1690$~MeV.

In other words, the models of Refs. \cite{skl,aniso} predict the
absence of any narrow
structures in the proton observables,
whereas the existence of the new N$^*$ should lead to the
presence of a narrow structure\footnote{In order to reveal such suppressed
narrow structures one has to consider observables with fine
binning in energy and with high enough statistics.} in observables for $\eta$ photoproduction on
the free proton.

One can put the two qualitatively different explanations of the
``neutron anomaly" to the test. If photoexcitation of a nucleon
resonance occurs on the neutron, its isospin partner must
materialize itself in the proton channel as well. Thus, {\it
experimentum crucis} lies in the studies of $\eta$ photoproduction
on the free proton. The observables in this case are not affected
by the nuclear effects.

In order to clarify the interpretation of the ``neutron anomaly"
we have undertaken in Ref.~\cite{KPB}\footnote{See this Ref. for
comments on the analysis of analogous data in Ref.\cite{ll}}
 a reanalysis of the GRAAL
data \cite{gra2,gra3} on the $\Sigma$ beam asymmetry for the
$\eta$ photoproduction off the free proton. We have extracted the
beam asymmetry using narrow energy bins, in order to reveal in
details the dependence of the beam asymmetry on the photon energy
in the region of $E_{\gamma}=0.85 - 1.15$~GeV (or $W=1.55 -
1.75$~GeV).

\begin{figure}
\vspace*{0.5cm}
\centerline{\resizebox{0.57\textwidth}{!}{\includegraphics{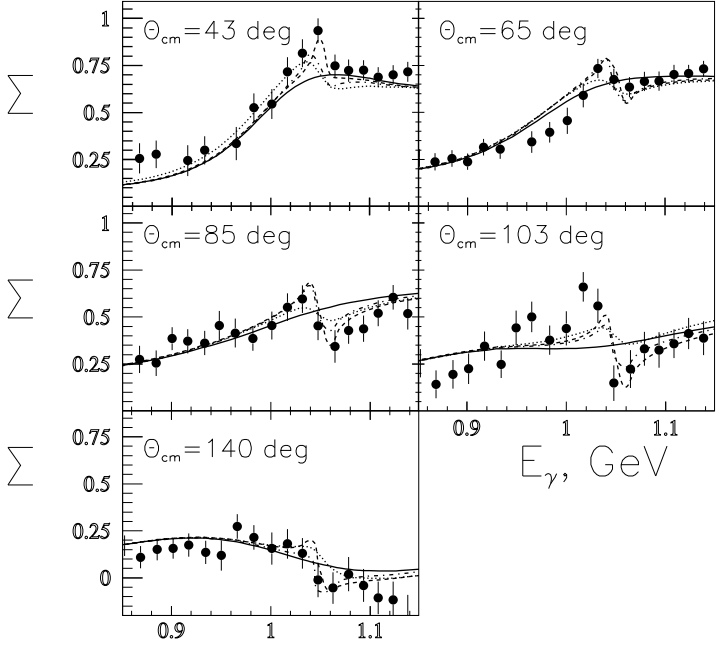}}}
\caption{Fig. 3. Fit of experimental data (filled circles data
obtained in the analysis of Ref.~\protect\cite{KPB}). Solid lines
show our calculations \protect\cite{KPB} based on the SAID
multipoles only, dotted lines include the $P_{11}$ resonance with
the width $\Gamma=19$~MeV; dashed lines are calculations with the
$P_{13}$ resonance ($\Gamma=8$~MeV), while the dash-dotted lines
use the resonance $D_{13}$, also with $\Gamma=8$~MeV.
 } \vspace*{-0.5cm} \label{fig:as2}
\vspace*{-0.3cm}
\end{figure}

The results of Ref.~\cite{KPB} are presented in
Fig.~\ref{fig:as2}. We see that the peak at forward angles and the
oscillating structure at central angles  form a pattern similar to
the interference of a narrow resonance with a smooth background.
In order to examine this assumption, we employ the multipoles of
the recent E429 solution of the SAID partial-wave
analysis~\cite{str1} for $\eta$ photoproduction as the model for
the smooth part of the observables. We see on Fig.~\ref{fig:as2}
that the SAID multipoles provide a good description of the data on
$\Sigma$ beam asymmetry. However, in the narrow photon energy
interval of $E_\gamma=1.015-1.095$ the considerable deviation of
the data from the smooth curve provided by the SAID multipoles takes
place. The $\chi^2$ value for the points in this energy interval
[$6\times 4$ points - 6 energy bins in 4 angular bins] for the
SAID solution is rather sizeable $\chi^2/dof=74/24$. For the
nearby energy bins the SAID solution gives a good description of the
data. A natural way to describe the deviation of the data from the
smooth SAID solution is an addition  of a
narrow resonance in the Breit-Wigner form (see e.g. \cite{maid}) to the SAID multipoles.
The contribution of a resonance is parameterized by the mass,
width, photocouplings (multiplied by square root of $\eta N$
branching), and the phase. These parameters are varied in order to
achieve the minimization of $\chi^2$.
\\

The mass of the included resonances is strongly constrained by the
experimental data. The mass values belong to the range of
$M_R=1.685 - 1.690$~GeV. The best agreement with the data is
obtained with the width of $\Gamma \sim 8$~MeV for $P_{13}$ and
$D_{13}$, and $\Gamma\sim 19$~MeV for $P_{11}$. However, the
reasonable reproduction of the data is achieved for the width up
to $\Gamma \leq 25$~MeV.

We tried various quantum numbers of the resonance. The $S_{11}$
resonance generates a dip at $43^{\circ}$ in the entire variation range of
 its photocoupling and its phase. It does not lead to
improvement of $\chi^2$ in the photon energy interval
$E_\gamma=1.015-1.095$. This indicates that most probably the observed
structures  can not be attributed to a narrow
$S_{11}$ resonance. The inclusion of the narrow $P_{11}, P_{13}$
and $D_{13}$ resonances improves the description of the data. The
corresponding values of $\chi^2$ are the following:
$\chi^2/dof=56/22$ for the $P_{11}$; $\chi^2/dof=25/20$ for the
$P_{13}$; and $\chi^2/dof=39/20$ for the $D_{13}$ resonances.

The curves shown in Fig.~\ref{fig:as2} correspond to the following
values of the photocouplings:
\begin{itemize}
\item
for the $P_{11}$ resonance:
\begin{equation}
\sqrt{Br_{\eta N}} A^p_{1/2}\sim 1\cdot 10^{-3}~{\rm GeV}^{-1/2};
\label{p11A12}
\end{equation}
\item
for the $P_{13}$ resonance:
\begin{eqnarray}
\sqrt{Br_{\eta N}} A^p_{1/2}&\sim& -0.3\cdot 10^{-3}{\rm ~GeV}^{-1/2},\\
\sqrt{Br_{\eta N}} A^p_{3/2}&\sim& 1.7\cdot 10^{-3}{\rm
~GeV}^{-1/2};
\end{eqnarray}
\item
for the $D_{13}$ resonance:
\begin{eqnarray}
\sqrt{Br_{\eta N}} A^p_{1/2}&\sim& -0.1\cdot 10^{-3}{\rm ~GeV}^{-1/2},\\
\sqrt{Br_{\eta N}} A^p_{3/2}&\sim& 0.9\cdot 10^{-3}{\rm
~GeV}^{-1/2}.
\end{eqnarray}
\end{itemize}
The neutron photocoupling is considerably larger than the above
values for the proton. On the base of the data from
Refs.~\cite{Kuznetsov04,gra1}, the photocoupling of the tentative
N$^*$ was estimated in Ref.~\cite{akps} as\footnote{Possible
theoretical errors of this analysis are up to a factor of two.}:
\begin{eqnarray}
\sqrt{{\rm Br}_{\eta N}} A_{1/2}^n \sim 15\cdot 10^{-3}~{\rm
GeV}^{-1/2}. \label{nA12}
\end{eqnarray}
The value (\ref{p11A12}) of $\sqrt{Br_{\eta N}} A^p_{1/2}$ and the
value (\ref{nA12}) of $\sqrt{{\rm Br}_{\eta N}} A_{1/2}^n$ are in
a good agreement with the estimates for the non-strange pentaquark
from the anti-decouplet performed in Chiral Quark-Soliton
Model~\cite{max,yang}. \\

In summary, we have demonstrated here that the existence of a new
narrow nucleon resonance N$^*$(1685) has sprouted from the
experimental results of Refs.~\cite{Kuznetsov04,gra1,kru,kas,KPB}.
 We have
 educed its properties from the data of
 Refs.~\cite{Kuznetsov04,gra1,kru,KPB} as follows:

 \begin{itemize}
 \item
 The mass is \cite{Kuznetsov04,gra1,kru,KPB}
$$M=1.685\pm 0.005\pm 0.007~{\rm GeV}.$$
\item
The width is estimated as \cite{KPB}
$$\Gamma \leq 25~{\rm MeV}.$$
\item
The neutron photocoupling is much stronger than that of the proton \cite{akps,KPB}
$$\frac{\Gamma(n^*\to n \gamma)}{\Gamma(p^*\to p \gamma)} \sim 50-250.$$
\item
Most probably the $S_{11}$ quantum numbers are excluded \cite{KPB}.
\end{itemize}
Employing additional information on elastic $\pi N$ scattering
\cite{str} and broken SU$_{\rm fl}$(3) \cite{guz2} we can obtain
further properties of N$^*$(1685):
\begin{itemize}
\item
The $\pi N$ branching is estimated as \cite{str}
$${\rm Br}_{\pi N}\leq 5\%.$$
\item
The most probable quantum numbers are $P_{11}$ \cite{str}.

\item
Mixing angles between N$^*(1685)$ and ground state nucleon,
$P_{11}(1440)$, and $P_{11}(1710)$ are small \cite{guz2}:
$$
|\theta_{1,2,3}|\leq 12^\circ\, .
$$
\end{itemize}
It seems that for many years we have been overlooking a {\it
narrow} nucleon resonance with a mass around 1685~MeV! Indeed,
searches for new baryon resonances have been focusing on the
states with a width in the range of hundreds of MeV. Such
``missing" resonances have been copiously predicted by
variants of the 3-quark models of baryons. The existence of an
excited nucleon state with a width of tens of MeV has been
unthinkable.

The new narrow nucleon N$^*$(1685) discussed here, being a very good
candidate for the non-strange member of the exotic anti-decouplet,
provides us with strong circumstantial evidence for the existence of  {\it Ultima Thule} of
hadronic physics -- the exotic $\Theta^+$ baryon.

There are several experiments that support the existence of
$\Theta^+$ baryon. We mention only two collaborations, that first
explored $\Theta^+$. The LEPS and DIANA collaborations not only
announced the pioneering signals in 2003  \cite{Nakano,Dolgolenko}
but also confirmed their signals on the higher statistics after a
careful and critical analysis \cite{diana2,leps2}. Yet, the
existence of $\Theta^+$ has not been widely accepted (see e.g.
\cite{Close}). One of the most influential negative results on
$\Theta^+$ comes from the report \cite{neg} by the CLAS
collaboration, in which the previous CLAS announcement \cite{clas}
of the evidence for $\Theta^+$ is renounced.
 The presented in Refs. \cite{clas} evidences for $\Theta^+$
were based on noncritical estimates of the background and
statistical significance of the announced signal, as it has been
shown by the most recent CLAS analysis\footnote{Note that even
this analysis is not flawless \cite{cousins} } \cite{Ireland}. The
reports of the CLAS collaboration in no way diminish the evidences
for $\Theta^+$ provided by the LEPS, DIANA and other
collaborations.

Firstly, the direct and indirect evidences for the existence of $\Theta^+$ are strong and can not be simply brushed away.

\section*{Acknowledgements}
We are grateful for discussions and support to M.~Amarian,
Ya.~I.~Azimov, D.~Diakonov, A.~G.~Dolgolenko, V.~Petrov,
M.~Praszalowicz and I.~Strakovsky. B.~Krusche is thanked for
providing us with Fig.~2. This work has been supported in part by
the Sofja Kowalewskaja Programme of Alexander von Humboldt
Foundation, by DFG (TR16), and in part by Korean Research
Foundation.


\end{document}